# Lithography-free sub-100nm nanocone array antireflection layer for low-cost silicon solar cell

**Zhida Xu[1*], Jing Jiang[1], Gang Logan Liu[1]**

[1]*Department of Electrical and Computer Engineering, University of Illinois at Urbana-Champaign, Urbana, IL 61801, USA*
[#]*Equally contributing authors*

* *zhidaxu1@illionis.edu*

High density and uniformity sub-100nm surface oxidized silicon nanocone forest structure is created and integrated onto the existing texturization microstructures on photovoltaic device surface by a one-step high throughput plasma enhanced texturization method. We suppressed the broadband optical reflection on chemically textured grade-B silicon solar cells for up to 70.25% through this nanomanufacturing method. The performance of the solar cell is improved with the short circuit current increased by 7.1%, fill factor increased by 7.0%, conversion efficiency increased by 14.66%. Our method demonstrates the potential to improve the photovoltaic device performance with low cost high and throughput nanomanufacturing technology.

## 1. Introduction

As a renewable and green energy source, solar energy is attracting increasing attentions nowadays but the relatively high cost and low efficiency of solar cells compared to fossil fuels still remain hurdles for their broader use. To overcome the hurdles, researchers are investigating new photovoltaic devices, such as III-V compound multi-junction solar cell with more than 37% efficiency [1, 2] , low cost printable polymer solar cell[3] and thin film solar cell[4]. Crystalline silicon(c-Si) solar cell still dominates the market considering the prevalence of silicon in electronic industry and the compromise of low cost, efficiency and simplicity in design and manufacturing [5]. As the Shockley–Queisser theory proposes, the maximum theoretical conversion efficiency of single junction silicon solar cell is 33.7%[6] however modern commercial crystalline solar cells produce only about 10%~20% conversion efficiency. The overall conversion efficiency is the product of optical absorption efficiency, thermodynamic efficiency, charge carrier separation efficiency and conductive efficiency[7]. The high optical reflectivity of the c-Si wafer surface is usually accounted for a major part in the total loss[8]. To suppress the optical reflection, techniques such as ZnO and $Si_3N_4$ antireflective coating(ARC)[9,10], nanowire antireflection[11], periodical patterning of surface[12,13] surface roughening and texturing using chemical wet etching[8,14,15], reactive ion etching(RIE)[14,15] and laser pulse irradiation[16-18]. Among these techniques, antireflection coating and surface periodical patterning only suppress the optical reflection for certain wavelength range and incident angle. The structures made by surface roughening and texturing are more random but can suppress reflection omindirectionally in broadband wavelength range. Alkaline pyramidal texturing is the most widely used antireflective texturing techniques in c-Si photovoltaic industry due to its convenience and low cost but it has several major disadvantages. Firstly, the anisotropic alkaline etching works best for monocrystalline silicon plane <100> while not good for other crystallographic planes or multicrystalline silicon. Secondly, the pyramidal structure created by alkaline is usually in micro-meter range thus not efficient for reflection suppression from visible to near-infrared range[14,21,22]. Other techniques such as laser pulse irradiation[16-18], self-mask RIE[14,19] and other chemical etching processes [20,21] can produce sub-wavelength structures on silicon to make it completely "black" with the optical reflection below 5%. However, most of these techniques are too complex and expensive to apply in industry and the nanostructure size is still in hundreds of nanometers range. There are two additional benefits if the surface texture structure can be made below 100 nm[8, 22]. Firstly, while deep surface texturing reduce the reflection mainly by multiple reflection and absorption, the fine features much smaller than wavelength reduce the reflection and enhance the transmission as effective medium with gradually varying optical refractive index[8, 22]. So the sub-100nm structure is more likely to transmit the light to the underneath p-n junction, where most of the photoelectrons are generated. According to Stephens and Cody's effective medium model, the reflectivity cutoff for textured silicon surfaces with a linearly increasing density should appear at a wavelength 4-6 times larger than the depth of the textured layer[23]. That means textures in sub-100nm depth ensure a proper subwavelength operation thus effectively enhances optical transmission. Secondly, since the doping depth of a solar cell is usually hundreds of nanometers, deep texturing may penetrate the doped layer to damage the p-n junction, harming the solar cell performance. That is why most of previous work performed doping after texturing of silicon rather than texture directly on the solar cell. Only with the texturing depth less than 100nm, we can even accomplish nanotexturing directly on finished solar cell products.

In this paper we demonstrated that with the one-step, wafer-scale, lithograph-free nanotexturization process, we are able to produce sub-100nm cone forest structures on pyramidal textured solar cell at room temperature and within a few minutes. The performance of the cheap solar cell is dramatically improved after treatment, with the open circuit voltage increased by 0.06%, short circuit current increased by 7.09%, fill factor increased by 7.0%, conversion efficiency increased by 14.66% and overall external quantum efficiency increased by 14.31%. This high throughput, low cost and mass-producible silicon surface nanotexturing process can be readily applied to current silicon photovoltaic devices.

## 2. Fabrication process of black silicon

The surface of silicon can be nanotexturized with the simultaneous plasma enhanced reactive ion etching and synthesis (SPERISE) process[24]

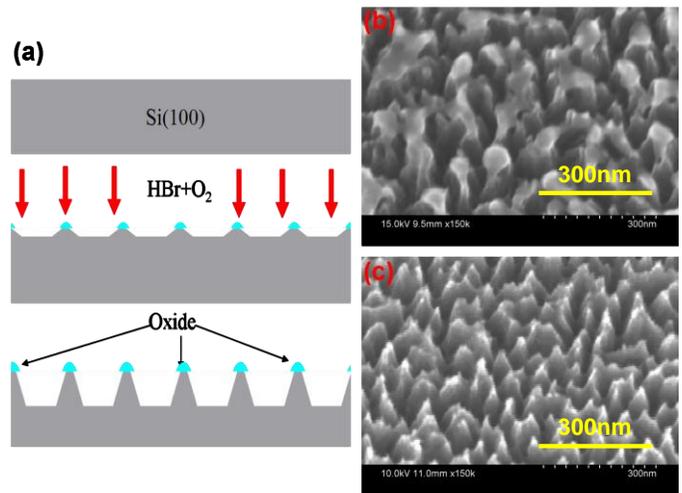

Fig. 1. (Color online) (a) SPERISE fabrication process of nanotexturized black silicon. (b) SEM image of black silicon surface before the removal of synthesized oxybromide nanoparticles. (c) SEM image of black silicon surface after the removal of oxybromide nanoparticles.

carried in a plasma etcher with $O_2$ and HBr gas mixture. The bromine ion plays the role of etching and texturizing while oxygen ion plays the role of nanosynthesis and oxidizing passivation. Figure 1 illustrates the fabrication process. $O_2$ with the flow rate of 15 sccm and HBr with the flow rate of 20 sccm are sent in simultaneously under the pressure of 100 mTorr. Radio frequency (RF) electromagnetic wave with the power of 200w and the frequency of 13.56 MHz is used to generate plasma. In the first few seconds a nanoparticle array made of silicon oxybromide naturally forms on silicon surface as the etch mask to protect the silicon underneath from being etched by HBr. In a few minutes, with the uncovered silicon being etched down the silicon underneath the nanoparticle masks ends up with a cone structure. We usually keep the plasma on for 7 minutes. For etching time more than 10 minutes, the nano cones can be etched away and disappear [25,29]. This one-step etching process can texturize the silicon surface with highly dense and uniform nanocone forest on the entire wafer surface. The oxybromide nanoparticles mask remains like a hat on top of the silicon cone when the etching process finished as shown in Fig. 1(b). After the oxybromide nanoparticle mask is removed with buffered oxide etch (BOE), the silicon nanocones look very sharp and dense as shown in Fig. 1(c). From the SEM pictures we can see that the silicon cones are about 80 nm in height, 40 nm in base width and 100 nm in spacing distance between adjacent cones.

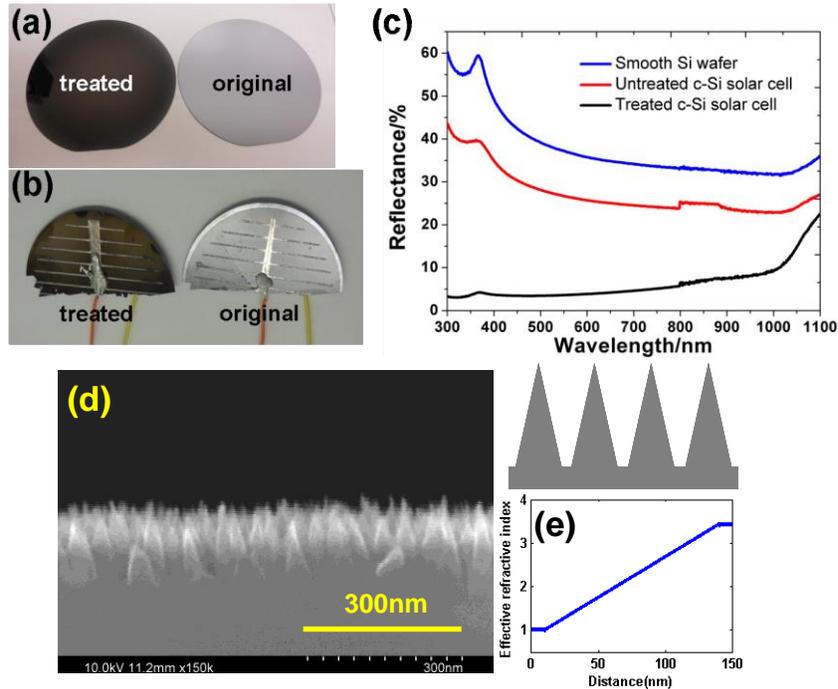

Fig. 2. (Color online) (a) Untreated silicon 3" <100> wafer (right) and Si wafer treated to be black silicon with the SPERISE method (left). (b) Untreated c-Si 2" solar cell (right) and nanotexturized c-Si solar cell with much darker and more diffusive surface (left). (c) Diffusive reflection spectra of smooth Si wafer (blue), untreated c-Si solar cell (red) and nanotexturized c-Si solar cell (black). (d) Cross-section SEM image of black silicon. (e) Schematic to show the effective refractive index of nanopillar array on black silicon

According to the effective media theory, the gradually varying refractive index of the cone forest structure can dramatically reduce the reflection thus makes the silicon surface black [8, 22] as shown in Fig. 2(a). So we call it black silicon. Fig. 2(d) shows the cross-sectional SEM image of the nanopillar and Fig. 2(e) is the schematic of gradually varying effective refractive index. With the reflection suppressed on the surface, it is possible to trap and covert more incident light to electrical energy so as to increase the external quantum efficiency of the solar cell. For black silicon produced with this SPERISE method, in addition to its application in photovoltaics, we also demonstrated elsewhere the implementation of its optical property in broadband surface-enhanced Raman scattering and fluorescence enhancement [25] and its applications in peptide sensing[26], detection of water contamination[27] and enhanced living cell imaging[28].

## 3.Nanotexturizing commercial solar cell with SPERISE

Our recipe of $O_2$-HBr plasma RIE can be applied to silicon surface with various morphologies. Figure 2(a) is the photograph to show the comparison of the appearance of RIE-treated plane silicon wafer <100> (left) and that of untreated plane silicon wafer (right). Conspicuously, by appearance the nanotexturized silicon wafer is black while the untreated one is bright and shiny. To demonstrate that our SPERISE method can improve the performance of commercial solar cells, we purchased grade-B silicon solar cells. Grade B means that the quality of this solar cell is low but it is cheap, cost 1 dollar per piece. The surface of the c-Si solar cell has already been textured with wet chemical etching so it looks somewhat greyish. However, the size of pyramidal textures is in tens of micrometers scale so it is not very efficient in broadband antireflection. But with the SPERISE process, we are able to make nanocone forest on the pyramidal microstructure to further reduce the reflection. To show that our method can change the appearance of solar cell, we cut one 3" commercial solar cell into two semicircles and treated one of them with our nanotexturization method. Figure 2(b) is the photograph to show the comparison of the appearance of the untreated commercial solar cell (right) and that of nanotexturized commercial solar cell (left). Obviously the nanotexturized solar cell is much "blacker" while the untreated solar cell is bright. Figure 3(a) and (b) are SEM images of the surface of the commercial solar cell and Fig. 3(c) and (d) are SEM images of the surface of the nanotexturized solar cell. The nanocone structure on pyramids looks similar to that on RIE-treated plane silicon wafer, in Fig. 1(c). Figure 2(c) is the reflectance spectra on untreated silicon wafer, untreated commercial solar cell and nanotexturized solar cell. With an integration sphere setup, we collected the diffusive reflection from all the angles and sent it to a spectrometer with the wavelength range from 300nm to 1100nm. The discontinuities at 800nm on all three spectra in Fig. 2(c) are due to switching the spectrometer detectors. The average reflectance can be defined as: [29]

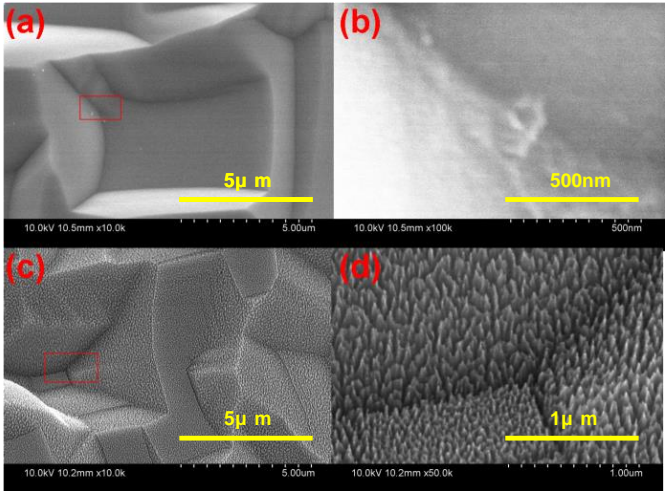

Fig. 3. (Color online) (a) SEM image of the surface of untreated c-Si solar cell. (b) Zoomed-in SEM image of the region on (a) indicated by the red square. (c) SEM image of the surface of nanotexturized c-Si solar cell. (d) Zoomed-in SEM image of the region on (c) indicated by the red square.

$$\frac{\int_{300}^{1100} R(\lambda)N(\lambda)d\lambda}{\int_{300}^{1100} N(\lambda)d\lambda} \quad (1)$$

where $R(\lambda)$ – total reflectance, $N(\lambda)$ – the solar flux under AM1.5 standard conditions. From Fig. 2(c) we know that the smooth silicon wafer has the highest reflectance in the whole wavelength range with the averaged reflectance of 45.2%; the nanotexturized solar cell has the lowest reflectance in the whole wavelength range with the average value of 9.4%; the original grade-B solar cell has the reflectance in between the above two in the whole wavelength range with the average value of 31.6%. So the SPERISE nanotexturization treatment can suppress the optical reflection by 70.25% for the commercial solar cell.

## 3. I-V characteristics and efficiency

To obtain the I-V characteristic curves and the conversion efficiencies of the commercial solar cell and the nanotexturized solar cell, we use a standard solar simulator. All the I-V curves are measured under the illumination of air mass (AM) 1.5 with the power density of 100 mW/cm$^2$ (one-sun) and at the temperature of 25 $^{\circ}$C. To ensure the uniform illumination and facilitate the calculation of conversion efficiency, during the measurement the solar cell is covered by a photomask with an opening window with the area of 1.6 cm$^2$. To eliminate the difference probably induced by different samples, we measure and compare the I-V curves of the same solar cell before and after the SPERISE treatment. Figure 4(a) shows the I-V curves of one solar cell before (red curve) and after (black curve) SPERISE treatment. We can see that after the SPERISE treatment, the I-V curve of solar cell shows improvement in parameters including the fill factor (FF), open circuit voltage ($V_{oc}$), short circuit current ($I_{sc}$) and maximum power ($P_{max}$). To figure out the conversion efficiency (η), we calculate the incident light power as 100 mW based on the illumination power density of 100 mW/cm$^2$ and the illuminated area (the area of the hole in the center of the photomask) of 1.6 cm$^2$. More quantitative data is listed in Table 1. After the solar cell is treated with the SPERISE method, the FF is increased from 47.49% to 50.81%, by 7.0%; the $V_{oc}$ stays almost the same value at 0.5V; the $I_{sc}$ is increased from 49.4% to 52.9%, by 7.09; the $P_{max}$ is increased from 11.6mW to 13.3mW, by 14.66% and most importantly, the conversion efficiency (η) is increased from 7.25% to 8.32%, by 14.66%. In Table 1, note that compared with the results in most previous literature about nanotexturization, the serial resistance ($R_s$) and shunt resistance ($R_{sh}$) of our solar cells did not change considerably after the SPERISE treatment. As we use commercial solar cell with metal contact already on it, when we treat the solar cell with the SPERISE method, we protected the metal contact with photoresist so the metal contact are not affected. This issue will be discussed in more details later in the discussion section.

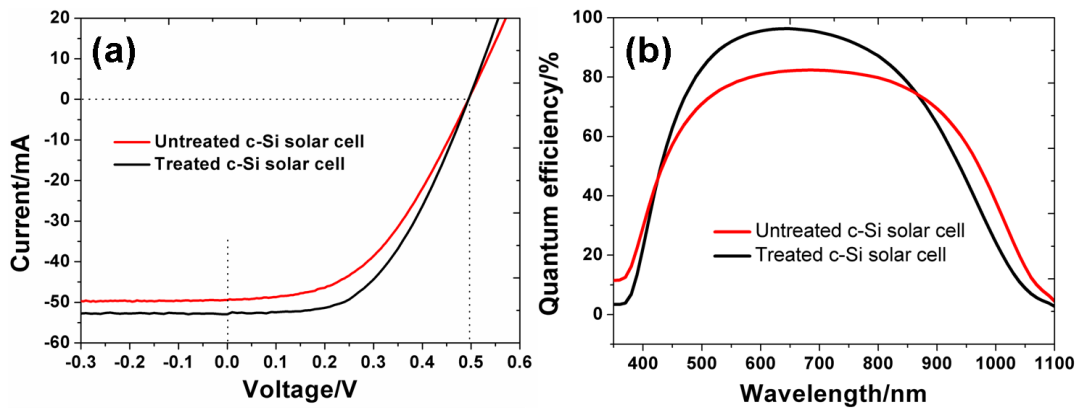

Fig. 4. (a) I-V characteristics of commercial solar cell before (red curve) and after (black curve) nanotexturization. The data were measured under the illumination of air mass (AM) 1.5 with the power density of 100 mW/cm$^2$ and at the temperature of 25 $^{\circ}$C. (b) External quantum efficiency spectra of commercial solar cell before (red curve) and after (black curve) SPERISE nanotexturization treatment.

To compare the quantum efficiency of solar cell before and after the SPERISE treatment at individual wavelength, we measured the external quantum efficiency (EQE) spectrum in the wavelength from 300 nm to 1100 nm. Figure 4(a) shows the EQE spectra of one solar cell before (red curve) and after (black curve) the SPERISE treatment. We can see that after the RIE- treatment even though the EQE is declined slightly in the ultraviolet (UV) range (300nm~400nm) and near infrared (NIR) range (850nm~1100nm), it is dramatically increased in most of the visible range (400nm~800nm). We define the average EQE similarly as using Eqn. 1 for the average reflectance. The averaged EQE of the solar cell in the wavelength range from 300nm to 800nm is 45.2% before the SPERISE treatment while 51.7% after the treatment, improved by 14.3%. This is what we desired since the solar spectrum is mostly concentrated in the visible range [30]. The enhanced EQE agrees with the enhanced conversion efficiency in the I-V measurement.

## 4. Discussion

We have demonstrated that after being nanotexturized by the SPERISE process for several minutes, the c-Si solar cell will be more efficient in converting light to electricity. However, for a more thorough understanding of why the efficiency gets improved and how to further improve it, several questions need to be answered.

First, while the diffusive reflection measurement shows the broadband optical absorption enhancement of 70.25% (Fig. 2(c) and Table 1), in comparison, the conversion efficiency measurement with I-V characterization is 14.66% and the average quantum efficiency enhancement is 14.31% (Fig. 4). The difference between the absorption enhancement and conversion efficiency indicates that the absorbed light is not fully converted to electricity. Obviously, the absorbed photons can go through different paths other than generating photoelectrons, such as generating heat and surface recombination. It is well known that the crystal defects and dangling bonds on silicon surface will result in the surface recombination thus deteriorate the photovoltaic conversion efficiency. If the planar silicon surface is textured, there is more chance for the existence of crystal defects and dangling bonds. So the surface texturization may decline the conversion efficiency to some extend. In terms of the conversion efficiency, there must be some competition between the increase by absorption enhancement and the decrease by surface recombination. That is one reason why the enhancement in conversion efficiency is lower than the absorption enhancement. But considering that the conversion efficiency was still dramatically enhanced by about 15%, we believe that the absorption enhancement dominates this competition.

Table 1. Comparison of the parameters of c-Si solar cell before and after the SPERISE nanotexturization treatment, including fill factor (FF), open circuit voltage ($V_{oc}$), short circuit current ($I_{sc}$), maximum output power ($P_{max}$), input power ($P_{in}$), conversion efficiency (Conv. Eff.), serial resistance ($R_s$) and shunt resistance ($R_{sh}$).

| Parameter | FF | Voc/V | Isc/mA | Pmax/mW | Pin/mW | Conv. Eff. | Rs/ohm | Rsh/ohm |
|---|---|---|---|---|---|---|---|---|
| Before | 47.49% | 0.4945 | 30.875 | 11.6 | 160 | 7.25% | 4.0027 | 262.99 |
| After | 50.81% | 0.4948 | 33.06 | 13.3 | 160 | 8.32% | 3.8373 | 278.08 |
| Improvement | 7.00% | 0.06% | 7.09% | 14.66% | | 14.66% | | |

Second, in Fig. 2 (c) we can see that the reflectance of solar cell surface is suppressed by the SPERISE treatment in the whole wavelength range from UV to IR. However, in Fig. 4 (b), the external quantum efficiency just got improved in most of the visible and near-IR region. This inconsistence can be explained with the reason previously mentioned. That is, the absorption enhancement dominates in the visible and near-IR region while the surface recombination dominants in other regions. But this is just a preliminary assumption. There may be multiple physical processes going on the nanotextured surface which need further investigations.

Third, why does the open circuit voltage ($V_{oc}$) stay the same as 0.5V after the solar cell is treated with the SPERISE process? This is predictable since the illumination of one sun is high. With the illumination increasing, $V_{oc}$ cannot go to the infinity but will be limited by the equilibrium contact potential[31].

Finally, what are the advantages of our method? We have already discussed some of them in the introduction section but it is necessary to summarize them here once again. The major point is that the nanostructures we created on the silicon solar cell surface are high uniformity and high density sub-100nm nanocone forest. With less than 100nm in size, the silicon nanocone is much smaller than those previously reported on black silicon. The benefit resulted from the sub-100nm geometry is that the finer nanostructure will work better as the effective media with smooth refractive index gradient to transmit light to the junction[8, 22]. The other good thing is the shallow nanotexturization will not go beyond the junction depth to damage the junction so we can directly nanotexturize the solar cell after doping and even after contact metal bonding. Texturization after doping and contact bonding can bring huge convenience in solar cell production and avoid two potential problems. One problem is that if the metal contact is created after the texturization the quality of the contact may be affected by the nanoscale topology. The other problem is that if the doping is done after the texturization the doping profile and doping depth may be affected by the surface structure and thus unpredictable. Some literatures reported that the conversion efficiency became declined when the solar cell is made by doping on black silicon and they attributed the decline to the non-uniform doping on black silicon[29]. Most importantly, our SPERISE method can be applied to any available solar cells with various surface morphologies. We have already demonstrated that we able to integrate the monolithically created nanostructures onto any existing silicon microstructures in Fig. 3.

## 5. Conclusion

With a one-step lithography-free RIE etching process, we are able to directly texturize the surface of commercial crystalline silicon solar cell with dense and uniform sub-100nm nanocone forest integrated on existing microstructures. The nanocone forest structure dramatically reduces the optical reflection by 70.25%. The I-V characteristics of the solar cell measured under one-sun indicates that our nanotexturization method can improve the open circuit voltage by a little, short circuit current by 7.09%, fill factor by 7.0%, and conversion efficiency by 14.66%. The quantum efficiency is also increased by 14.31%. This rapid, low-cost and high-throughput nanotexturization method has the immediate potential to improve the efficiency and bring down the manufacturing cost of photovoltaic devices.